%% file: paper.tex
\documentclass[cernpreprint,texlive=2011,txfonts,UKenglish,texmf]{atlasdoc}
\pdfoutput=1

\usepackage{atlaspackage}
\usepackage{siunitx}
\usepackage{rotating}
\usepackage[rightcaption]{sidecap}

\usepackage{atlasphysics}

\hypersetup{pdftitle={ATLAS draft},pdfauthor={The ATLAS Collaboration}}

\AtlasTitle{Measurement of the Inelastic Proton-Proton Cross Section at $\sqrt{s} = 13 \TeV$ with the ATLAS Detector at the LHC}

\PreprintIdNumber{CERN-EP-2016-140}
\AtlasJournalRef{Phys.\ Rev.\ Lett. 117 (2016) 182002}
\AtlasDOI{10.1103/PhysRevLett.117.182002}

\AtlasAbstract{
This Letter presents a measurement of the inelastic proton-proton cross section using 60~\SI{}{\micro\barn}$^{-1}$ of $pp$ collisions at a center-of-mass energy $\sqrt{s}$ of $13$~\TeV\ with the ATLAS detector at the LHC. Inelastic interactions are selected using rings of plastic scintillators in the forward region ($2.07<|\eta|<3.86$) of the detector.  A cross section of $68.1\pm 1.4$~mb is measured in the fiducial region $\xi=M_X^2/s>10^{-6}$, where $M_X$ is the larger invariant mass of the two hadronic systems separated by the largest rapidity gap in the event.  In this $\xi$ range the scintillators are highly efficient.  For diffractive events this corresponds to cases where at least one proton dissociates to a system with $M_X>13$~\GeV. The measured cross section is compared with a range of theoretical predictions. When extrapolated to the full phase space, a cross-section of $78.1 \pm 2.9~\mbox{ mb}$ is measured, consistent with the inelastic cross section increasing with center-of-mass energy.}

\begin{document}

\maketitle

%-------------------------------------------------------------------------------
The rise of the total proton-proton ($pp$) cross section with center-of-mass energy $\sqrt{s}$, predicted by Heisenberg~\cite{Heisenberg} and observed at the CERN Intersecting Storage Rings~\cite{ISR}, probes the non-perturbative regime of quantum chromodynamics (QCD). Arguments based on unitarity, analyticity, and factorization imply an upper bound on the high-energy behavior of total hadronic cross sections that prevents them from rising more rapidly than ln$^2(s)$~\cite{froissart,martin,martin-inel}. 

Many experiments have measured $\sigma_\text{inel}$ and found an increase with $\sqrt{s}$~\cite{otherexp}. 
The TOTEM and ATLAS collaborations determined $\sigma_\text{inel}$ at $\sqrt{s}=7$ and 8~\TeV\ using the optical theorem and a measurement of the elastic cross section with Roman pot detectors~\cite{totem,STDM-2013-10,totem8,totem2,alfa8}. Using a variety of  alternative techniques, the ATLAS, CMS, ALICE, and LHCb experiments have made measurements of $\sigma_\text{inel}$ at $\sqrt{s}=7$~\TeV~\cite{STDM-2010-11,CMS-FWD-11-001,alice,lhcb} and $\sqrt{s}=2.76$~\TeV (ALICE)~\cite{alice}. 
The Pierre Auger Collaboration measured the inelastic $p$-air cross section at $\sqrt{s}=57$~\TeV\ and extracted $\sigma_\text{inel}$ using the Glauber model~\cite{auger}.

This Letter presents a measurement of the inelastic cross section $\sigma_\text{inel}$ using $pp$ collisions at $\sqrt{s}=13$~\TeV\ with the ATLAS detector at the Large Hadron Collider~(LHC). It is performed using two sets of scintillation counters in a dataset corresponding to an integrated luminosity of $60.1 \pm 1.1$~\SI{}{\micro\barn}$^{-1}$ collected in June 2015. In inelastic interactions, one or both protons dissociate as a result of colored (non-diffractive) or colorless (diffractive) exchange.  The counters are insensitive to elastic $pp$ scattering and diffractive dissociation processes in which neither proton dissociates into a system, $X$, of mass $M_X>13$~\GeV, or equivalently, $\xi=M_X^2 / s>10^{-6}$.  The cross-section measurement is reported in this fiducial region, $\xi>10^{-6}$, and after extrapolation to the total inelastic cross section using models of inelastic interactions. 

The ATLAS detector is a cylindrical particle detector\footnote{ATLAS uses a right-handed coordinate system with its origin at the nominal interaction point (IP) in the center of the detector and the $z$-axis along the beam pipe. The $x$-axis points from the IP to the center of the LHC ring, and the $y$-axis points upward. Cylindrical coordinates $(r,\phi)$ are used in the transverse plane, $\phi$ being the azimuthal angle around the beam pipe. The pseudorapidity is defined in terms of the polar angle $\theta$ as $\eta=-\ln\tan(\theta/2)$.} composed of several subdetector layers~\cite{PERF-2007-01}.
The inner tracking detector (ID) is immersed in a 2~T magnetic field provided by a superconducting solenoid.
Around the tracker is a system of electromagnetic and hadronic calorimeters, which use liquid argon and lead, copper, or tungsten absorber for the electromagnetic and forward ($|\eta|>1.7$) hadronic components of the detector, and scintillator-tile active material and steel absorber for the central ($|\eta|<1.7$) hadronic component.

At $z=\pm 3.6$~m, thin plastic scintillation counters, the minimum-bias trigger scintillators (MBTS), are installed on the front face of each endcap calorimeter.  These detectors cover the region $2.07<|\eta|<3.86$.  They are similar to those described in Ref.~\cite{PERF-2007-01} but were rebuilt during 2014, when the coverage was slightly extended from $2.08<|\eta|<3.75$ after the $\sqrt{s}=7$~\TeV run.  
The MBTS are divided into inner (4 counters in $149<r<445$~mm) and outer (8 counters in $444.5<r<895$~mm) octagonal rings.

The ATLAS experiment uses a multi-stage trigger to select events at about 1~kHz for offline analysis.  Three trigger configurations were used to collect data for this analysis.  The primary triggers use the MBTS detector and constant-fraction discriminators to select events when two proton bunches collide in the detector.  To facilitate background studies, data were also collected with the same selection when no proton bunch (``empty'') or a single proton bunch from only one of the two beams (``single beam'') was passing through the center of ATLAS.  All of these triggers require at least one MBTS hit above threshold. Two additional triggers were used to collect data to determine the MBTS trigger efficiency, requiring either hits in a forward ($5.6 < |\eta| < 5.9$) Cerenkov detector (LUCID) or a far forward ($|\eta| > 8.4$) tungsten-scintillator calorimeter detector (LHCf~\cite{LHCf}) located at $z=\pm 17$~m and $\pm 140$~m, respectively.
The LHCf detector is an independent detector, but for the runs considered in this analysis, its trigger signals were incorporated into the ATLAS readout.  

Monte Carlo (MC) simulation samples were produced to correct the fiducial measurement and to compare to the data.  The detector response is modeled using a simulation based on {\sc Geant4}~\cite{Geant4-1,Geant4-2,SOFT-2010-01}.  The data and MC simulated events are passed through the same reconstruction and analysis software.  

The primary MC samples are based on the {\sc Pythia8} generator~\cite{Pythia8-1,Pythia8-2} either with the A2~\cite{ATL-PHYS-PUB-2012-003} set of tuned underlying-event parameters and the MSTW 2008 LO PDF set~\cite{MSTW} or with the Monash~\cite{MonashTune} set of tuned parameters and the NNPDF 2.3 LO PDF set~\cite{NNPDF}.  The samples are divided into four components: single-dissociation (SD, $pp\rightarrow pX$), double-dissociation (DD, $pp\rightarrow X Y$), central-dissociation (CD, $pp\rightarrow p X p$), all involving colorless exchange, and non-diffractive dissociation (ND) wherein color flow is present between the two colliding protons.  For all dissociation event types, the Monash tune is used.

{\sc Pythia8} uses a pomeron-based diffraction model~\cite{IngelmanSchlein} to describe colorless exchange with a default pomeron flux model by Schuler and Sj\"{o}strand (SS)~\cite{SchulerPomeron,PythiaDiffraction}. Alternative MC samples are generated with the pomeron flux model of Donnachie and Landshoff (DL)~\cite{DLPomeron} and with the minimum-bias Rockefeller (MBR) model~\cite{MBR}. In the DL model, the pomeron Regge trajectory is given by $\alpha(t) = 1 + \varepsilon + \alpha' t$, where $\varepsilon$ and $\alpha'$ are free parameters. In most samples used for this analysis, the value of $\alpha'$ is 0.25, the {\sc Pythia8} default. The $\varepsilon$ parameter is varied from 0.06 to 0.10 (the {\sc Pythia8} default is 0.085).  An additional sample produced with $\alpha'=0.35$ is found to be statistically consistent with the $\alpha'=0.25$ default samples in each aspect of this analysis. The ranges of $\varepsilon$ and $\alpha'$ considered are motivated by previous total, inelastic, elastic, and diffractive cross-section measurements, including measurements of low-mass diffraction by the ATLAS and CMS collaborations~\cite{STDM-2011-01,CMS-FSQ-12-005}. For the DL and SS models the CD component is neglected. 
The MBR model is tuned to data as described in Ref.~\cite{MBR} and includes a small CD component.  

The {\sc Epos LHC} and {\sc QGSJet-II} event generators are also used to simulate $pp$ collisions.  {\sc Epos LHC}~\cite{Epos} uses a ``cut pomeron'' model for diffraction and differs significantly from {\sc Pythia8} in its modeling of hadronization and the underlying event. {\sc QGSJet-II}~\cite{QGSJet,QGSJet2} uses Reggeon field theory to describe pomeron-pomeron interactions. Both {\sc Epos LHC} and {\sc QGSJet-II} have been developed primarily to model cosmic-ray showering in the atmosphere. 

The fiducial region of the measurement is determined using MC simulation.  In each generated event, the largest rapidity gap between any two final-state hadrons is used to define the boundary between two collections of hadrons.  These collections define the dissociation systems in an event-generator-independent manner.
The invariant mass of each collection is calculated, and the larger of the two masses, denoted $M_X$, is used to define $\xi=M_X^2 / s$. The variable $\xi$ is constrained to be $>6\times 10^{-9}$ by the elastic limit of $m_p^2 / s$ where $m_p$ is the proton mass.  This measurement is restricted to $\xi>10^{-6}$,  the region in which the event selection efficiency exceeds 50\%.

Two samples of data events passing the MBTS trigger requirements are selected: an inclusive sample and a single-sided sample.  The inclusive selection requires at least two MBTS counters with a charge above 0.15~pC ($n_{\rm MBTS} \geq 2$).  This threshold is chosen to be well above the electronic noise level of the counters. Requiring two hits rather than one substantially reduces background due to collision-induced radiation and activation. To constrain the diffractive component of the cross section and reduce the uncertainty in extrapolation to $\sigma_\text{inel}$, an additional single-sided selection is defined, requiring hits in at least two counters on one side of the detector and no hits on the other.  In the data, 4,159,074 events pass the inclusive selection and 442,192 events pass the single-sided selection.

The fiducial cross section is determined by

\begin{equation}
\sigma_\text{inel}^\text{fid} \left( \xi > 10^{-6} \right) = \frac{ N - N_\text{BG} } {\epsilon_{\text{trig}} \times \cal{L} } \times \frac {1 - f_{\xi<10^{-6} } } { \epsilon_\text{sel} },
\label{eq:main}
\end{equation}

\noindent where $N$ is the number of observed events passing the inclusive selection, $N_\text{BG}$ is the number of background events, $\epsilon_{\text{trig}}$ and $\epsilon_\text{sel}$ are factors accounting for the trigger and event selection efficiencies, $1-f_{\xi<10^{-6}}$ accounts for the migration of events with $\xi < 10^{-6}$ into the fiducial region, and $\cal{L}$ is the integrated luminosity of the sample.

Sources of background include interactions between the beam and residual gas in the beam pipe; interactions between the beam and collimators upstream of the detector, which can send charged particles through the detector parallel to the beam; collision-induced radiation; and activation backgrounds.  Backgrounds from cosmic rays and instrumental noise are negligible.    The mean number of $pp$ collisions in the same LHC bunch crossing was $2.3\times 10^{-3}$ for the recorded dataset. Thus, the contribution from multiple collisions are also negligible.  The beam-related background components are extracted from single-beam events and dominate the total background. They are normalized by scaling the number of selected single-beam events by a factor of $37/4\times 2$, accounting for the 37 colliding pairs of bunches and 4 bunches producing the single-beam data in this run.  The factor of $2$ accounts for the presence of two colliding bunches. The number of protons per bunch producing these single-beam events agrees with that in the colliding bunches to within 10\%. The radiation and activation-induced backgrounds are implicitly part of this background estimate. Double-counting of these components is removed using estimates from empty events. The total background contributions to the inclusive and single-sided data samples are determined to be 1.2\% and 5.8\% respectively. The classification of single-sided events as double-sided due to noise or other backgrounds is estimated to be below 0.1\%. A systematic uncertainty of 50\% is assigned to the background based on studies of the background composition and the relative contributions of the background components. This uncertainty is treated as fully correlated between the single-sided and inclusive selections.

The trigger efficiency for events passing the inclusive selection, $\epsilon_{\text{trig}}$, is measured with respect to events selected with the LUCID detector after subtracting the background.
A trigger efficiency of 99.7\% (97.4\%) is measured for the inclusive (single-sided) event sample. In both cases the statistical uncertainty is below 0.1\%. The efficiency is also measured with events selected by the LHCf detector and agrees within $\pm 0.3\%$ with the LUCID determination.  This difference is taken as a systematic uncertainty.

The ratio of the number of events passing the single-sided event selection to the number passing the inclusive selection ($R_\text{SS}$) is used to adjust, for each model, the fractional contribution of the single- and double-diffractive dissociative cross section ($\sigma_{\rm SD}+\sigma_{\rm DD}$) to the inelastic cross section, $f_{\rm D}=(\sigma_{\rm SD}+\sigma_{\rm DD})/\sigma_{\rm inel}$~\cite{STDM-2010-11}.   The measured value is $R_\text{SS}=10.4\%$ with a total uncertainty of $ \pm 0.4\%$. The dominant systematic uncertainty arises from the background subtraction in the single-sided sample. For each MC model, $f_{\rm D}$ is varied until it matches the observed $R_\text{SS}$ value in data. The data uncertainty is used to set the error in the constrained $f_{\rm D}$ for each model. An additional uncertainty in the ratio of single- to double-diffractive events is determined by taking the diffractive events to be entirely SD or to be evenly divided between SD and DD.

\begin{figure}[h]
\begin{center}
\includegraphics[width=0.68\textwidth]{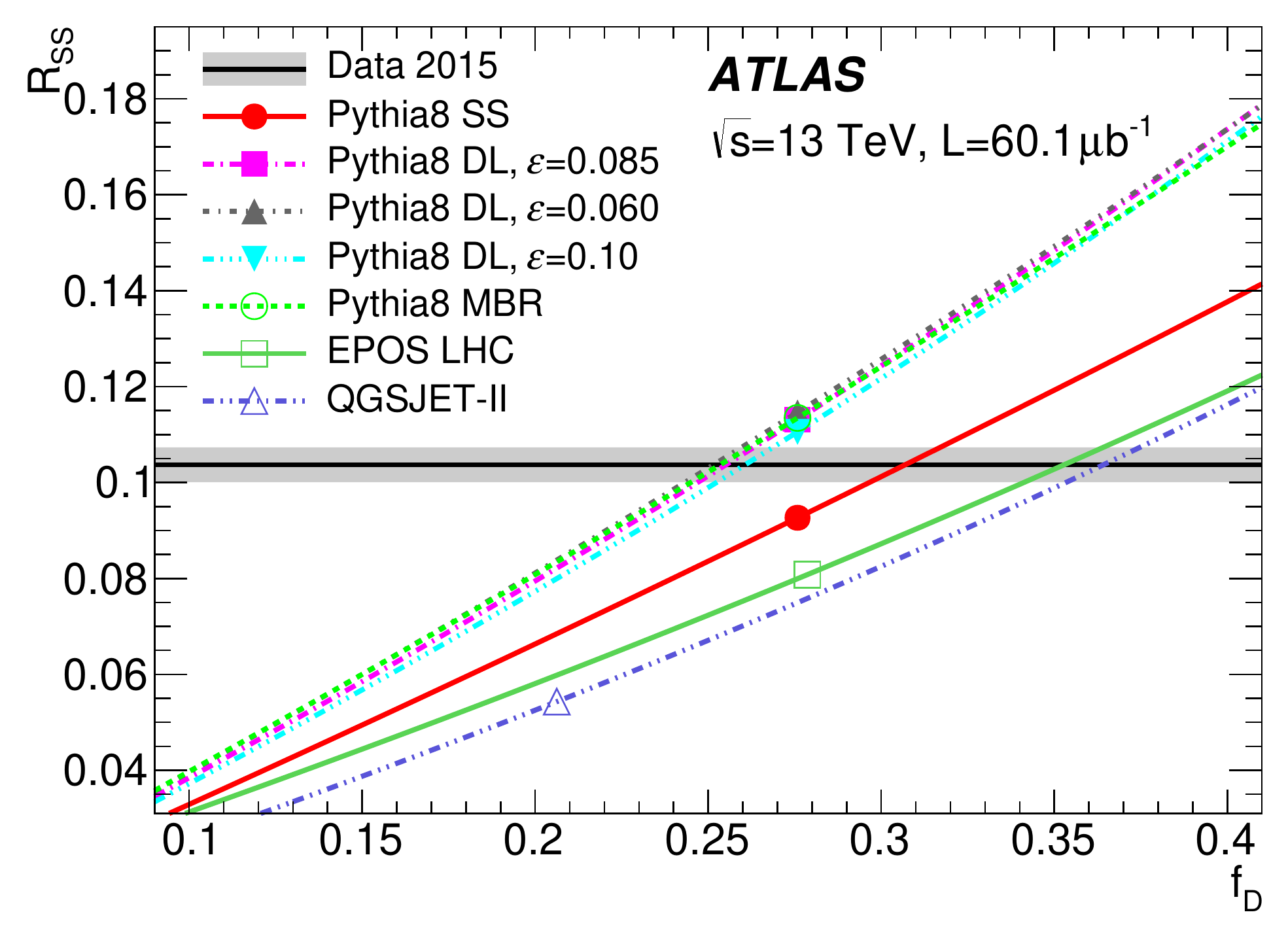}
\caption{The ratio of the number of single-sided to inclusive events ($R_\text{SS}$) as a function of the fraction of the cross section that is diffractive according to each model ($f_{\rm D}$).  The default value of $f_{\rm D}$ in each model is shown with a marker.}
\label{fig:fDRSS}
\end{center}
\end{figure}

Using this method,  the fitted $f_{\rm D}$ in the {\sc Pythia8} samples is between 25\% and 31\%, depending on the model (the default value is 28\%). For the {\sc QGSJet-II} ({\sc Epos LHC}) model the fitted $f_{\rm D}$ is  35\% (37\%), differing significantly from the default value of 21\% (28\%). The observed $R_\text{SS}$ and the MC predictions of its dependence on $f_{\rm D}$ are shown in Figure~\ref{fig:fDRSS}. The fitted $f_{\rm D}$ is used when determining the acceptance corrections $\epsilon_\text{sel}$ and $f_{\xi < 10^{-6}}$ for each model.

\begin{figure}
\begin{center}
\includegraphics[width=0.8\textwidth]{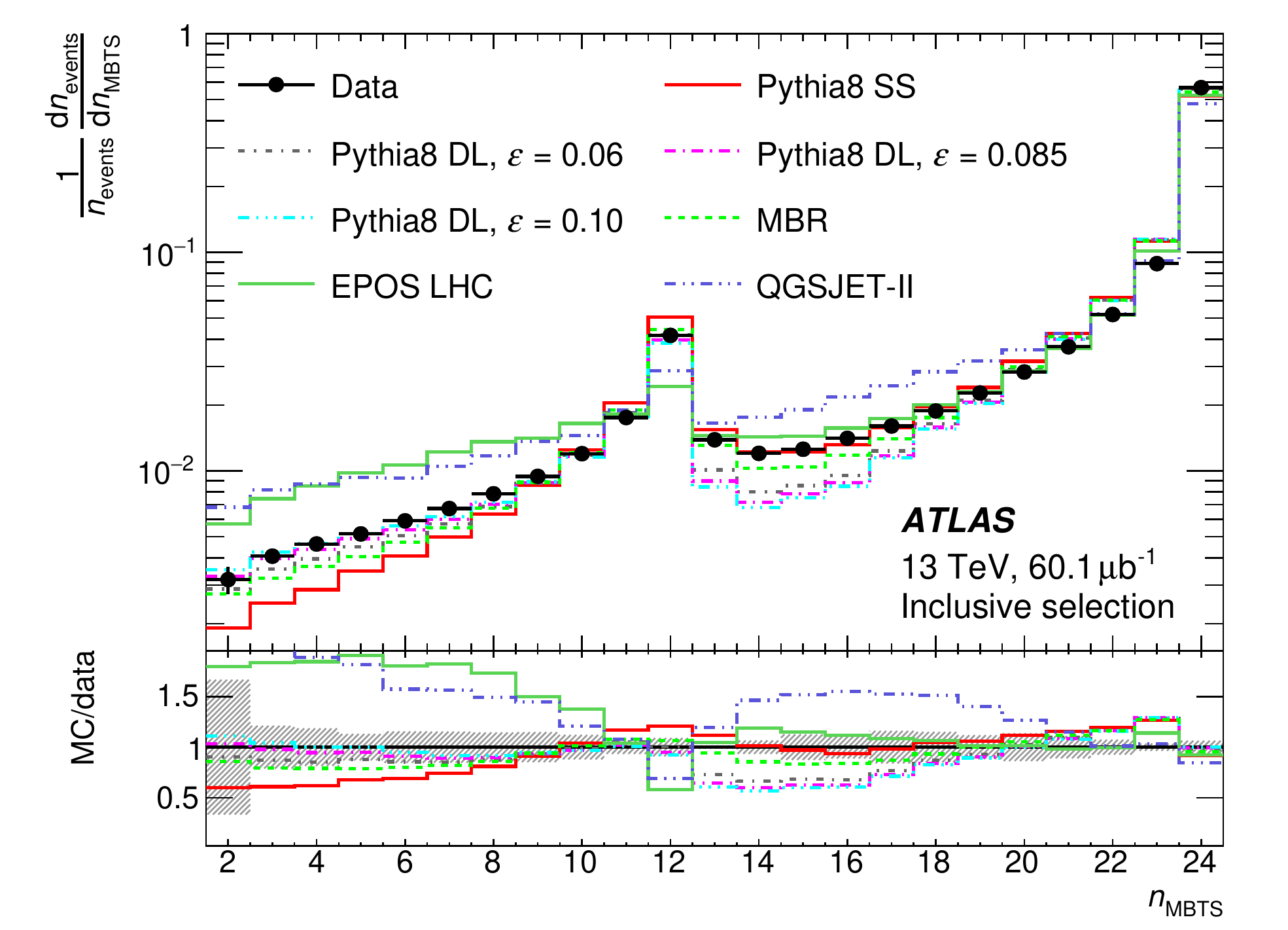}\\
\includegraphics[width=0.8\textwidth]{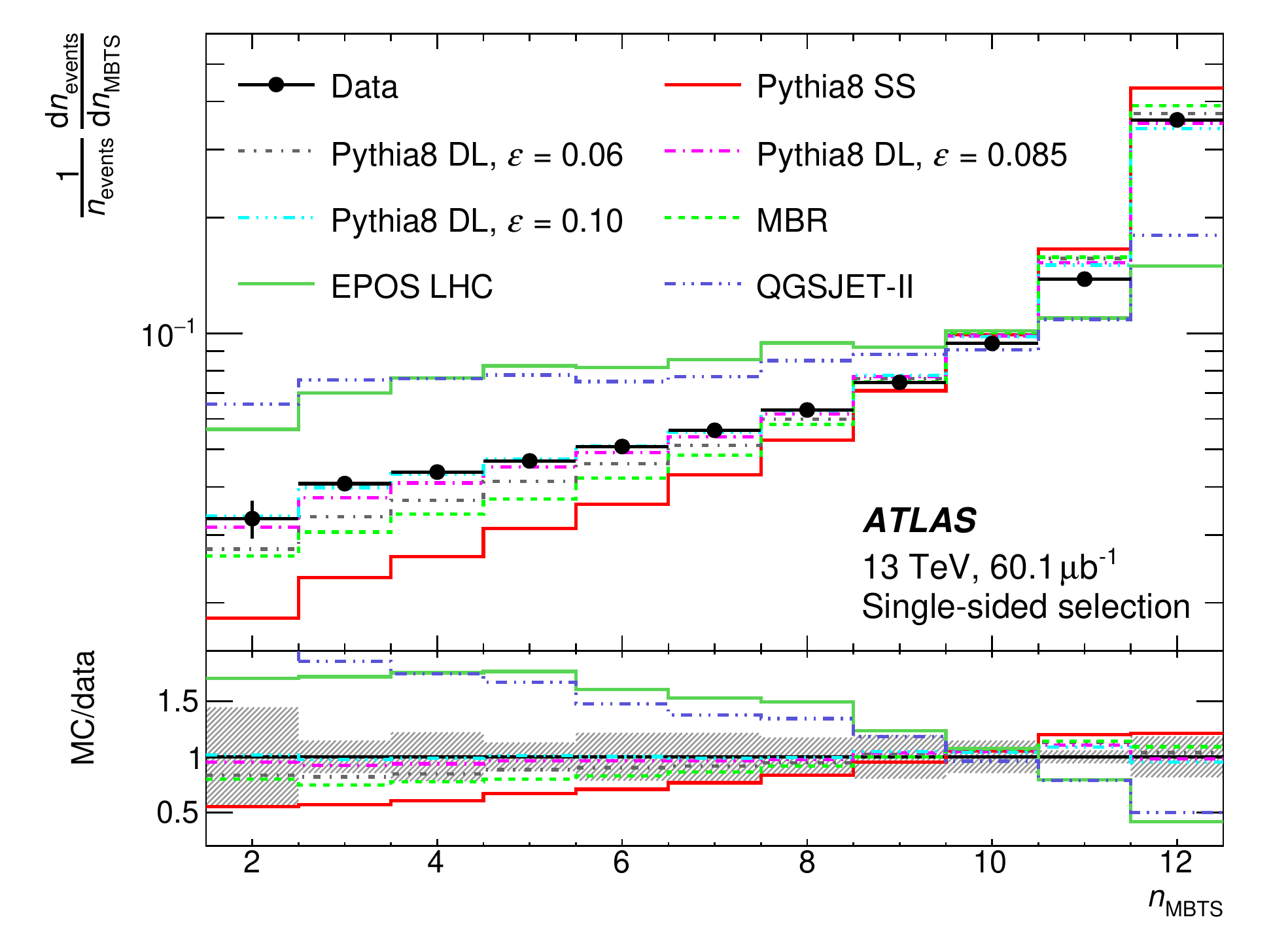}
\caption{
The background-subtracted distribution of the number of MBTS counters ($n_\text{MBTS}$) above threshold in data and MC simulation for (top) the inclusive selection and (bottom) the single-sided selection. The ratio of the MC models to the data is also shown. The experimental uncertainty is shown as a shaded band around the data points. The models shown here use the $f_D$ value determined from the $R_\text{SS}$ measurement.
}
\label{fig:mbts_counters}
\end{center}
\end{figure}

In Figure~\ref{fig:mbts_counters} the $n_{\rm MBTS}$ distributions in data are compared to the ones from MC simulated samples utilizing the fitted $f_{\rm D}$ values for both the inclusive and single-sided selections. The estimated background is subtracted from the measured distribution, and the trigger efficiency measured in data is applied to the simulation. The data distributions and MC simulation are peaked at high multiplicity values. In the single-sided case, $n_{\rm MBTS}=12$ corresponds to hits in all counters on one side of the detector. The data agree best with the DL models, particularly in the low-$n_{\rm MBTS}$ range. The MBR-based distribution provides a slightly worse description of the data. The {\sc Pythia8} sample using the SS model does not describe data well in the low-multiplicity region. {\sc Epos LHC} and {\sc QGSJet-II} also do not describe the data well, particularly in the single-sided hit multiplicity distribution. Therefore, the {\sc Pythia8} DL model with $\varepsilon=0.085$ is chosen as the nominal MC model for the $\epsilon_{\rm sel}$ and $f_{\xi < 10^{-6}}$ corrections, and only the DL and MBR models are considered for systematic uncertainties related to the MC corrections.

The event selection efficiency, $\epsilon_{\text{sel}}$, depends upon the MBTS counter sensitivity.  This sensitivity is tested using isolated charged particles, reconstructed as ID tracks in the region $2.07<|\eta|<2.5$ where the coverages of the MBTS and ID overlap.  Over the full coverage of the MBTS counters, the calorimeter is used to measure the counter efficiency with respect to particles that deposit sufficient energy in the calorimeter to seed a topological energy cluster~\cite{JetInputs}.  Differences between the efficiencies in data and MC simulation are accounted for by adjusting the MBTS charge threshold in MC simulation until the simulated efficiencies match those observed in the data. The residual uncertainty in the counter efficiency after these corrections is $\pm 0.5\%$ for the outer and $\pm 1.0\%$ for the inner counters. Additionally, an uncertainty arising from the knowledge of the material in front of the MBTS detector is estimated using MC samples with an increased amount of material in front of the MBTS. Based on the MC samples, the uncertainty in the efficiency measurement due to modeling of hadronization and the underlying event is estimated to be negligible.

After adjusting the counter charge threshold, $\epsilon_{\rm sel}$ is determined from the nominal {\sc Pythia8} DL MC, using the fitted $f_{\rm D}$ corresponding to this model, to be 99.34\% with a statistical uncertainty of $\pm$0.03\%. The uncertainty in the MBTS counter efficiencies results in only a $\pm$0.1\% uncertainty in the overall event selection efficiency, because many counters are hit in typical events. In addition, an uncertainty of $\pm 0.2\%$ in $\epsilon_{\rm sel}$ arises from the knowledge of the material in front of the MBTS.

The fraction of events passing the inclusive selection with $\xi<10^{-6}$ represents an additional background component in the fiducial cross-section measurement.  It is determined using the same {\sc Pythia8} DL MC to be $f_{\xi<10^{-6}}=(1.37\pm 0.05)$\%, where the uncertainty is statistical.

Because the efficiency and migration corrections are correlated, they are combined in a single correction factor, $C_{\rm MC}= (1 - f_{\xi<10^{-6} } ) / \epsilon_{\text{sel}}$, for which systematic uncertainties are assessed. The systematic uncertainties include the counter efficiency variations, the impact of the material uncertainty, the uncertainty in the fitted value of $f_{\rm D}$, and the variation in $C_{\rm MC}$ found by comparing the {\sc Pythia8} DL and MBR models.  Of these sources of uncertainty, the last is most important at $\pm$0.5\%.  The value of $C_{\rm MC}$ is ($99.3 \pm 0.5$)\%. The uncertainty also implicitly contains an uncertainty due to the CD contribution, since this is included in only some of the models.

The uncertainty in the integrated luminosity is $\pm1.9$\%. It is derived, following a methodology similar to that detailed in Refs.~\cite{DAPR-2011-01,Lumi2}, 
from a calibration of the luminosity scale using $x$--$y$ beam-separation scans performed in August 2015. This calibration uncertainty is slightly smaller than what has been reported in Ref.~\cite{WZxsec} because the low-luminosity dataset used in this letter is not affected by the uncertainties related to high-luminosity runs.

The components of the fiducial cross-section calculation [Eq.~\eqref{eq:main}] are shown in Table~\ref{tab:resultsingredients} with their systematic uncertainties. The statistical uncertainties are negligible.  The measured fiducial cross section is determined to be 

\begin{equation}
\sigma_\text{inel}^\text{fid} = 68.1\pm0.6\ \mbox{(exp.)}\ \pm1.3\ \mbox{(lum.)\ mb}, \nonumber
\end{equation}

where the first uncertainty refers to all experimental uncertainties apart from the luminosity and the second refers to the luminosity only. 

\begin{table}
\begin{center}
\caption{Inputs to the calculation of the measured cross section and their systematic uncertainties.  \label{tab:resultsingredients}}
\begin{tabular}{|l|r|r|}
\hline
Factor &  Value & Rel. uncertainty \\
\hline
Number of events passing the inclusive selection ($N$)	& 4159074	& $-$ \\
Number of background events ($N_\text{BG}$)       		& 51187		& $\pm 50\%$ \\
Integrated luminosity [\SI{}{\micro\barn}$^{-1}$] ($\cal{L}$)	& 60.1		& $\pm 1.9\%$ \\
Trigger efficiency ($\epsilon_{\text{trig}}$)       			& $99.7\%$	& $\pm 0.3\%$ \\
MC correction factor ($C_{\mathrm{MC}}$)                   	& $99.3\%$	& $\pm 0.5\%$ \\
\hline
\end{tabular}
\end{center}
\end{table}

The {\sc Pythia8} DL model predicts values of $71.0$~mb, $69.1$~mb and 68.1~mb for $\varepsilon=0.06$, 0.085 and 0.10 respectively,  all of which are compatible with the measurement. The {\sc Pythia8} MBR model predicts $70.1$~mb, also in agreement with the measurement. The {\sc Epos LHC} (71.2~mb) and {\sc QGSJet-II} (72.7~mb) predictions exceed the data by 2--3$\sigma$. The {\sc Pythia8} SS model predicts 74.4~mb, and thus exceeds the measured value by $\sim4\sigma$.

The extrapolation to $\sigma_\text{inel}$ uses constraints from previous ATLAS measurements to minimize the model dependence of the component that falls outside the fiducial region. $\sigma_\text{inel}$ can be written as

\begin{equation}\label{eqn:xsec}
\sigma_\text{inel} = \sigma_\text{inel}^\text{fid} + \sigma^\text{7~\TeV}(\xi<5\times10^{-6}) \times \frac{ \sigma^\text{MC}( \xi<10^{-6} ) } {\sigma^\text{7~\TeV, MC}(\xi<5\times10^{-6}) }.
\end{equation}

The term $\sigma^\text{7~\TeV}(\xi<5\times10^{-6})=\sigma^\text{7~\TeV}_\text{inel}-\sigma^\text{7~\TeV}(\xi>5\times10^{-6})=9.9\pm 2.4$~mb is the difference between $\sigma_\text{inel}$ measured at 7~\TeV\ using the ALFA detector~\cite{STDM-2013-10}, $\sigma_\text{inel}^\text{7~\TeV}$, and $\sigma_\text{inel}$ measured at 7~\TeV\ for $\xi>5\times 10^{-6}$ using the MBTS~\cite{STDM-2010-11}.\footnote{The 7~\TeV\ result is corrected upward by 1.9\% following an improved luminosity calibration~\cite{DAPR-2011-01}.} The uncertainties of the two measurements are uncorrelated.

The {\sc Pythia8} DL and {\sc Pythia8} MBR MC samples are used to assess the systematic uncertainty in the MC-derived ratio of cross sections in Eq.~(\ref{eqn:xsec}), which is determined to be $1.015 \pm0.081$\footnote{The value of the ratio arises from an approximately 20\% increased cross section from increasing $\sqrt{s}$ which is largely compensated by a 15\% decrease due to the change in the $\xi$ distribution.}. These models also agree with the measurement of $\sigma^\text{7~\TeV}(\xi<5\times10^{-6})$ to within $2\sigma$.

The measured value for $\sigma_\text{inel}$ is

\begin{equation}
\sigma_\text{inel} = 78.1 \pm 0.6~\mbox{(exp.)}\ \pm1.3~\mbox{(lum.)}\ \pm2.6~\mbox{(extrap.)\ mb.} \nonumber
\end{equation}

\begin{figure}
\begin{center}
\includegraphics[width=0.8\textwidth]{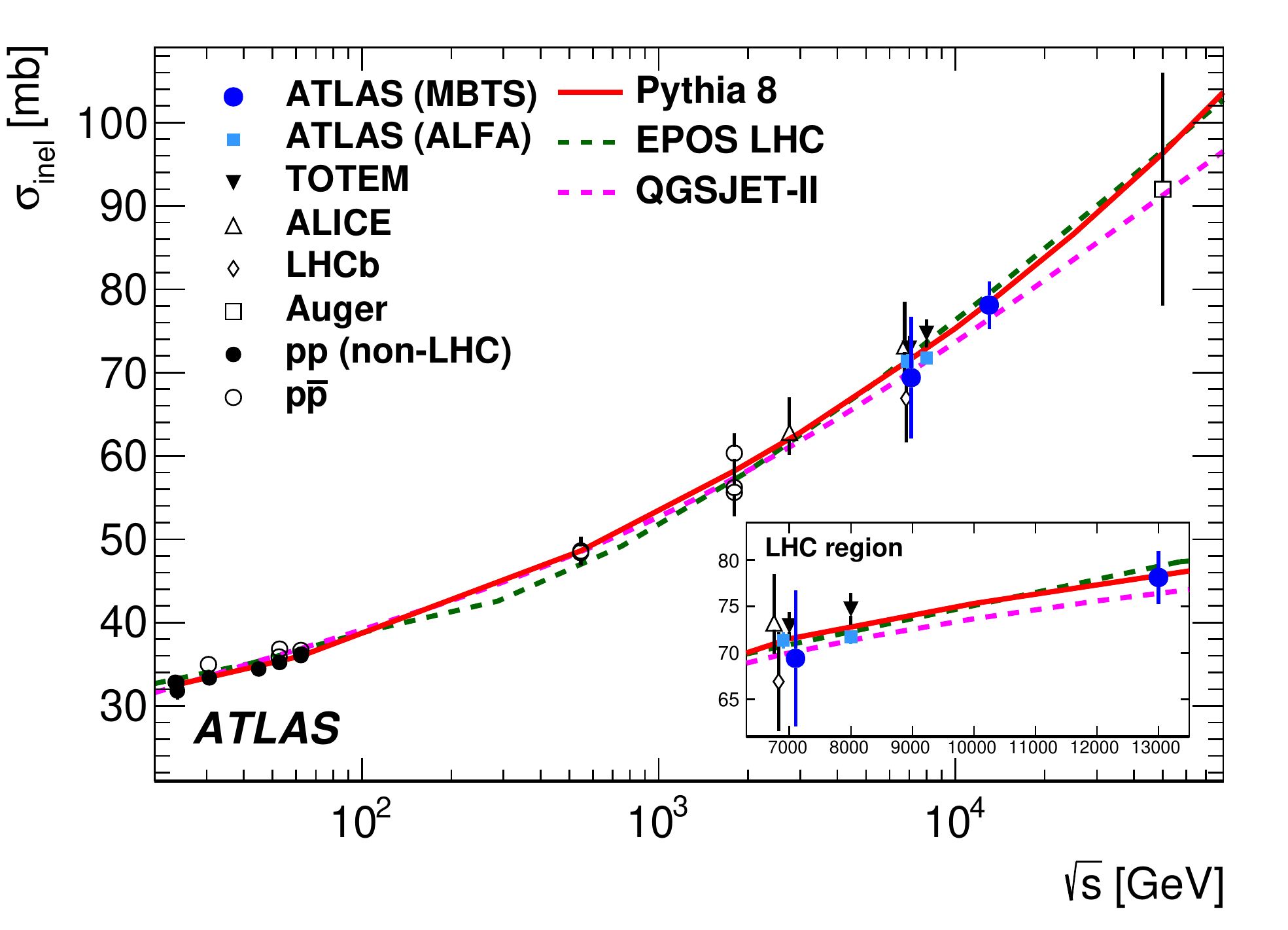}
\caption{The inelastic proton-proton cross section versus $\sqrt{s}$.
Measurements from other hadron collider experiments~\cite{otherexp,totem8,totem,lhcb,alice} and the Pierre Auger experiment~\cite{auger} are also shown. Some LHC data points have been slightly shifted in the horizontal position for display purposes. The data are compared to the {\sc Pythia8}, {\sc Epos LHC} and {\sc QGSJet-II} MC generator predictions.
The uncertainty in the ATLAS ALFA measurement is smaller than the marker size.}
\label{fig:predictions}
\end{center}
\end{figure}

\noindent This and other inelastic cross-section measurements are compared to several Monte Carlo models in Figure~\ref{fig:predictions}. 
Additional predictions range between 76.6 and 81.6~mb~\cite{1208.5446,1212.5096,1312.3851,1408.3811,1504.04890}.
Compared to the measurement with the ALFA detector at $\sqrt{s}=7$~TeV the cross section is higher by $(9 \pm 4)$\%.

In summary, a measurement of the inelastic cross section in 60~\SI{}{\micro\barn}$^{-1}$ of proton-proton collision data at ${\sqrt{s}=13}$~\TeV\ collected with the ATLAS detector at the LHC is presented.  The measurement is performed in a fiducial region $\xi>10^{-6}$, and the result is extrapolated to the inelastic cross section using measurements at $\sqrt{s}=7$~TeV. The measured cross section agrees well with a variety of theoretical predictions and is consistent with the inelastic cross section increasing with center-of-mass energy, as observed at lower energies. 

\FloatBarrier

%-------------------------------------------------------------------------------
\section*{Acknowledgements}
%-------------------------------------------------------------------------------

We acknowledge and thank the LHCf Collaboration for the use of their
triggers to check the MBTS trigger efficiency for this analysis.

% Acknowledgements for papers with collision data
% Version 2-Aug-2016

% Standard acknowledgements start here
%----------------------------------------------
We thank CERN for the very successful operation of the LHC, as well as the
support staff from our institutions without whom ATLAS could not be
operated efficiently.

We acknowledge the support of ANPCyT, Argentina; YerPhI, Armenia; ARC, Australia; BMWFW and FWF, Austria; ANAS, Azerbaijan; SSTC, Belarus; CNPq and FAPESP, Brazil; NSERC, NRC and CFI, Canada; CERN; CONICYT, Chile; CAS, MOST and NSFC, China; COLCIENCIAS, Colombia; MSMT CR, MPO CR and VSC CR, Czech Republic; DNRF and DNSRC, Denmark; IN2P3-CNRS, CEA-DSM/IRFU, France; GNSF, Georgia; BMBF, HGF, and MPG, Germany; GSRT, Greece; RGC, Hong Kong SAR, China; ISF, I-CORE and Benoziyo Center, Israel; INFN, Italy; MEXT and JSPS, Japan; CNRST, Morocco; FOM and NWO, Netherlands; RCN, Norway; MNiSW and NCN, Poland; FCT, Portugal; MNE/IFA, Romania; MES of Russia and NRC KI, Russian Federation; JINR; MESTD, Serbia; MSSR, Slovakia; ARRS and MIZ\v{S}, Slovenia; DST/NRF, South Africa; MINECO, Spain; SRC and Wallenberg Foundation, Sweden; SERI, SNSF and Cantons of Bern and Geneva, Switzerland; MOST, Taiwan; TAEK, Turkey; STFC, United Kingdom; DOE and NSF, United States of America. In addition, individual groups and members have received support from BCKDF, the Canada Council, CANARIE, CRC, Compute Canada, FQRNT, and the Ontario Innovation Trust, Canada; EPLANET, ERC, FP7, Horizon 2020 and Marie Sk{\l}odowska-Curie Actions, European Union; Investissements d'Avenir Labex and Idex, ANR, R{\'e}gion Auvergne and Fondation Partager le Savoir, France; DFG and AvH Foundation, Germany; Herakleitos, Thales and Aristeia programmes co-financed by EU-ESF and the Greek NSRF; BSF, GIF and Minerva, Israel; BRF, Norway; Generalitat de Catalunya, Generalitat Valenciana, Spain; the Royal Society and Leverhulme Trust, United Kingdom.

The crucial computing support from all WLCG partners is acknowledged gratefully, in particular from CERN, the ATLAS Tier-1 facilities at TRIUMF (Canada), NDGF (Denmark, Norway, Sweden), CC-IN2P3 (France), KIT/GridKA (Germany), INFN-CNAF (Italy), NL-T1 (Netherlands), PIC (Spain), ASGC (Taiwan), RAL (UK) and BNL (USA), the Tier-2 facilities worldwide and large non-WLCG resource providers. Major contributors of computing resources are listed in Ref.~\cite{ATL-GEN-PUB-2016-002}.
%----------------------------------------------

%-------------------------------------------------------------------------------
\printbibliography
%-------------------------------------------------------------------------------

\newpage \input{atlas_authlist}

\end{document}

%% file: atlas_authlist.tex
% ATLAS Collaboration author list
% Data extracted on 26-May-2016 for paper reference STDM-2015-05
\begin{flushleft}
{\Large The ATLAS Collaboration}

\bigskip

M.~Aaboud$^\textrm{\scriptsize 135d}$,
G.~Aad$^\textrm{\scriptsize 86}$,
B.~Abbott$^\textrm{\scriptsize 113}$,
J.~Abdallah$^\textrm{\scriptsize 64}$,
O.~Abdinov$^\textrm{\scriptsize 12}$,
B.~Abeloos$^\textrm{\scriptsize 117}$,
R.~Aben$^\textrm{\scriptsize 107}$,
O.S.~AbouZeid$^\textrm{\scriptsize 137}$,
N.L.~Abraham$^\textrm{\scriptsize 149}$,
H.~Abramowicz$^\textrm{\scriptsize 153}$
% ...
\bigskip
\\
% ...
$^{12}$ Institute of Physics, Azerbaijan Academy of Sciences, Baku, Azerbaijan\\
% ...
$^{64}$ University of Iowa, Iowa City IA, United States of America\\
% ...
$^{86}$ CPPM, Aix-Marseille Universit{\'e} and CNRS/IN2P3, Marseille, France\\
% ...
$^{107}$ Nikhef National Institute for Subatomic Physics and University of Amsterdam, Amsterdam, Netherlands\\
% ...
$^{113}$ Homer L. Dodge Department of Physics and Astronomy, University of Oklahoma, Norman OK, United States of America\\
% ...
$^{117}$ LAL, Univ. Paris-Sud, CNRS/IN2P3, Universit{\'e} Paris-Saclay, Orsay, France\\
% ...
$^{135}$ $^{(a)}$ Facult{\'e} des Sciences Ain Chock, R{\'e}seau Universitaire de Physique des Hautes Energies - Universit{\'e} Hassan II, Casablanca; $^{(b)}$ Centre National de l'Energie des Sciences Techniques Nucleaires, Rabat; $^{(c)}$ Facult{\'e} des Sciences Semlalia, Universit{\'e} Cadi Ayyad, LPHEA-Marrakech; $^{(d)}$ Facult{\'e} des Sciences, Universit{\'e} Mohamed Premier and LPTPM, Oujda; $^{(e)}$ Facult{\'e} des sciences, Universit{\'e} Mohammed V, Rabat, Morocco\\
% ...
$^{137}$ Santa Cruz Institute for Particle Physics, University of California Santa Cruz, Santa Cruz CA, United States of America\\
% ...
$^{149}$ Department of Physics and Astronomy, University of Sussex, Brighton, United Kingdom\\
% ...
$^{153}$ Raymond and Beverly Sackler School of Physics and Astronomy, Tel Aviv University, Tel Aviv, Israel\\
% ...
$^{a}$ Also at Department of Physics, King's College London, London, United Kingdom\\
$^{b}$ Also at Institute of Physics, Azerbaijan Academy of Sciences, Baku, Azerbaijan\\
$^{c}$ Also at Novosibirsk State University, Novosibirsk, Russia\\
$^{d}$ Also at TRIUMF, Vancouver BC, Canada\\
$^{e}$ Also at Department of Physics {\&} Astronomy, University of Louisville, Louisville, KY, United States of America\\
% ...
\end{flushleft}

% Created with xml2latex.py